\documentclass[11pt,a4paper]{article}

\usepackage[T1]{fontenc}
\usepackage[utf8]{inputenc}
\usepackage{lmodern}
\usepackage{titlesec}
\usepackage[top=2.5cm, bottom=2.5cm, left=2.5cm, right=2.5cm]{geometry}

\usepackage{amsmath}
\usepackage{amssymb}
\usepackage{amsthm}
\usepackage{longtable}  
\usepackage{graphicx}
\usepackage{float}
\usepackage{tabularx}
\usepackage{booktabs}
\usepackage{multirow}
\usepackage{caption}
\usepackage{subcaption}
\usepackage{adjustbox}

\usepackage{xcolor}
\usepackage[colorlinks=true,
            linkcolor=blue,
            citecolor=blue,
            urlcolor=blue]{hyperref}
\usepackage{doi}

\usepackage{gensymb}
\usepackage{placeins}
\usepackage{lipsum}  

\usepackage{setspace}
\usepackage{authblk}

\usepackage{abstract}

\titleformat{\section}
  {\normalsize\bfseries}{\thesection}{1em}{}

\titleformat{\subsection}
  {\small\bfseries}{\thesubsection}{1em}{}

\titleformat{\subsubsection}
  {\footnotesize\bfseries}{\thesubsubsection}{1em}{}
\begin{document}

\title{\normalsize\textbf{Integrating Artificial Neural Networks into Undergraduate Physics Laboratory: A Compound Pendulum Case Study}}

\author[1*]{\small Saralasrita Mohanty}
\author[1]{\small Prabhu Prasad Tripathy}
\author[2]{\small Raja Das}
\author[1]{\small Sudakshina Prusty}

\affil[1]{School of Physical Sciences, National Institute of Science Education and Research, Jatni 752050, India}
\affil[2]{Department of Mathematics, School of Advanced Sciences, Vellore Institute of Technology, Vellore, 632 014, Tamil Nadu, India}

\date{\today}

\maketitle

\begin{center}
    \small\textit{*Corresponding author: \href{mailto:saralasrita@niser.ac.in}{saralasrita@niser.ac.in}}
\end{center}

\begin{abstract}

Artificial Neural Networks (ANNs) have emerged as powerful tools in physics research and education by enabling data-driven modeling and complementing traditional analytical approaches. Incorporating ANNs into undergraduate physics laboratories can provide students with exposure to modern computational techniques alongside conventional experimental analysis. In this study, ANN implementation is introduced in a traditional compound pendulum experiment for determining the acceleration due to gravity, $g$. The objective is not to replace the analytical method, but to demonstrate how machine-learning-based models can be integrated into experimental physics as complementary tools for data analysis, regression, and model validation. Students first measure experimentally relevant parameters such as effective length, time period, and initial angular displacement, and determine $g$ using standard analytical methods with associated experimental uncertainty. These experimentally obtained parameters are then used as inputs to an ANN model consisting of input, hidden, and output layers. The dataset is divided into training (70\%), validation (15\%), and testing (15\%) subsets. The ANN model reproduces the experimentally observed relationship between the pendulum parameters and $g$ with high consistency across the training, validation, and test datasets. The experimentally determined value of gravitational acceleration was $1009.03 \pm 6.82~\mathrm{cm/s^2}$, while the ANN model produced a mean predicted value of $1009.029858~\mathrm{cm/s^2}$ with a mean absolute prediction error of $0.000592~\mathrm{cm/s^2}$. The close agreement between the experimental and ANN-predicted values indicates that the neural network successfully learned the relationship between the measured pendulum parameters and $g$. However, the ANN prediction error should not be interpreted as an improvement in the intrinsic experimental precision, since the model is trained using experimentally derived quantities related through the known analytical equation. Rather, the ANN serves as a complementary computational and pedagogical tool that introduces students to concepts such as regression, validation, overfitting, and data-driven analysis alongside traditional experimental physics.

\end{abstract}

\section{Introduction} 
The pendulum remains a classic example in physics, demonstrating principles of inertia, gravity and oscillation \cite{Ref1}, \cite{Ref2}, \cite{Ref3}, \cite{Ref4}. Specifically, in the undergraduate laboratories, the compound pendulum experiment is treated as a pedagogical activity to study rigid body dynamics, the concept of moment of inertia and to determine the local value of (g) \cite{Ref5}, \cite{Ref6}, \cite{Ref7}, \cite{Ref8}, \cite{Ref9}.Conventional methods for measuring ‘g’ depend on theoretical models that presume ideal conditions such as a uniform rigid-body pendulum, frictionless pivots, negligible air resistance and small oscillations. However, real experiments deviate from ideal assumptions since experimental observations are influenced by factors such as noise, damping, and non-linear effects due to finite amplitude \cite{Ref10}. 
Undergraduate laboratories traditionally introduce students to experimental uncertainty analysis involving both random and systematic errors \cite{Ref11}, \cite{Ref12}, \cite{Ref13}, \cite{Ref14}, \cite{Ref15}, \cite{Ref16}, \cite{Ref17}. Random errors arise from unpredictable variations such as human reaction time during timing measurements, whereas systematic errors may originate from calibration inaccuracies, frictional effects, or environmental influences. Although these analytical methods form an essential part of physics education, recent advances in computational techniques have created opportunities for integrating data-driven approaches into experimental analysis \cite{Ref18}.
Artificial Neural Networks (ANNs), a subset of machine learning, have emerged as useful computational tools for recognizing patterns, approximating nonlinear relationships, and analyzing experimental datasets in physics and engineering \cite{Ref19}, \cite{Ref20},\cite{Ref21}. Their growing applications in scientific research and data analysis have also motivated interest in introducing such techniques into physics education \cite{Ref22}, \cite{Ref23}, \cite{Ref24}. In undergraduate laboratories, ANN-based approaches can help students explore concepts such as regression, prediction, training, validation, and model generalization using experimentally obtained data.

The increasing integration of computational and data-driven methodologies into modern scientific research has made familiarity with machine learning, numerical analysis, and scientific programming increasingly important for physics students. Incorporating ANN-based analysis into familiar undergraduate experiments provides an accessible framework through which these computational concepts can be introduced without losing connection to established physical intuition. By embedding machine learning within a well-known experimental context such as the compound pendulum, students can engage with modern computational tools in a conceptually grounded manner. This approach lowers the barrier to interdisciplinary learning while demonstrating how contemporary data-driven techniques can complement traditional experimental and analytical methods in physics.

In this work, ANN modeling is incorporated into the traditional compound pendulum experiment as a complementary computational framework for analyzing experimentally obtained data. Students first determine \(g\) using standard analytical procedures and experimental measurements, after which the measured parameters are used to train and evaluate an ANN model. The ANN functions as a model-independent approximator that reconstructs the underlying oscillatory behaviour from experimental measurements without explicitly imposing the analytical form of harmonic motion. Importantly, the ANN is not presented as a means of improving the intrinsic accuracy of the experiment; rather, it serves as a pedagogical tool for illustrating how data-driven modeling differs from conventional analytical fitting techniques.

By comparing traditional statistical analysis with ANN-based reconstruction, students are introduced to two complementary paradigms of scientific modeling: physics-driven and data-driven approaches. This comparison encourages critical discussion of noise, uncertainty, interpolation, physical interpretability, and the limitations of machine learning in experimental science. The integration of ANN modeling into undergraduate laboratories therefore provides a contemporary educational framework that connects classical experimental physics with emerging computational practices widely used in modern scientific research.

\begin{figure}[H]
  \centering
  \includegraphics[width=0.8\textwidth]{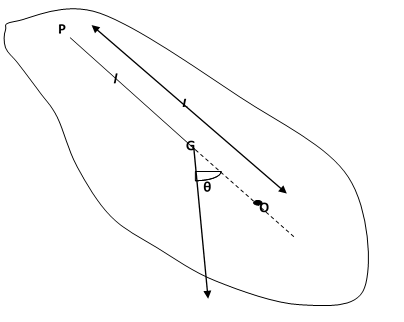}
  \caption{Oscillation of a Compound Pendulum}
  \label{fig:single}
\end{figure}

\section{Theoretical Framework and Analytical Formulation of the Compound Pendulum}
A compound pendulum, also known as a physical pendulum, consists of a rigid body that swings about a fixed horizontal axis under the influence of gravity. Unlike the simple pendulum, where all the mass is assumed to be concentrated at a single point, the compound pendulum takes into account the distribution of mass and its moment of inertia about the pivot point. Consider a rigid body of mass m that is freely suspended from a horizontal frictionless axis. Let the center of gravity (G) be located at a distance l from the pivot (P), as illustrated in Figure 1. If the pendulum is displaced slightly from its equilibrium position, the equation of motion for the compound pendulum is expressed as:
\begin{equation}
    \frac{d^2\theta}{dt^2} + \frac{mgl}{I\theta} = 0  
\end{equation}
where $\theta$ is the angular displacement and I is the moment of inertia of the pendulum about the point of suspension, P. This is a standard equation for simple harmonic motion, where the time period of oscillation, T,  can be expressed as:
\begin{equation}
    T = \frac{2\pi}{\omega} = 2\pi \sqrt{\frac{I}{mgl}}
\end{equation} 
where $\omega$ is the angular frequency. The moment of inertia I about the pivot point can be determined using the parallel axis theorem:
\begin{equation}
  I = mK^2 + ml^2  
\end{equation}
where K is the radius of gyration about the axis passing through G. Substituting Eq. 3 in Eq.2 The expression for the time period can be rewritten as:
\begin{equation}
    T = 2\pi \sqrt{\frac{(K^2 + l^2 )}{lg}}
\end{equation}
This result indicates that the time period of a compound pendulum is equivalent to that of a simple pendulum with an effective length $L_{eff}$, which can be measured as the distance between the point of suspension P and the centre of oscillation O in Figure 1 such that $T = 2\pi \sqrt{\frac{L_{eff}}{g}}$, where $L_{eff}$  can be expressed as follows:
\begin{equation}
    L_{eff} = l + \frac{K^2}{l} 
\end{equation}

This equation shows that the effective length is a combination of the distance l from the pivot to the center of gravity and an additional term that accounts for the mass distribution\cite{Ref25}. Also, solving the quadratic equation expressed in Eq. 5 one can obtain two solutions $l_1$ and $l_2$ such that
\begin{equation}
    l_1+ l_2 = L_{eff}\ and\ l_1l_2 = K^2 
\end{equation}	  
Since $K^2$ is positive, thus both $l_1$ and $l_2$ are positive. This means that on either side of the centre of gravity there are two positions of the centre of suspension about which the time periods are the same. Thus there are four such positions on the pendulum with the same time period. The distance between two such positions of the centers of suspension, asymmetrically located on either side of the centre of gravity, can be experimentally estimated as $L_{eff}$  as per Eq. 6.
The compound pendulum experimental set up consists of a rigid body that rotates about a fixed axis through a fixed point. In our experiment, a 100 cm metal bar was used with pre-drilled holes positioned at equal distances of 5cm along its length. A digital timer was employed to measure the time taken by the metal bar to complete 10 oscillations. This timing was recorded for five repetitions at each hole on the bar. To satisfy small oscillation conditions, the initial angular amplitude was varied between 1-5 degrees. By plotting the time period T versus the distance l, one can estimate the effective length graphically using Eq.6. Then the ‘g’ can be determined using the following expression:
\begin{equation}
    g = \frac{4\pi^2L_{eff}}{T^2}  
\end{equation}
 	
This forms the basis for comparing experimental results with analytical predictions, further validating the theoretical framework.

\section{Methodology for developing the ANN model and the experimental setup}
\subsection{Understanding the ANN architecture}
An Artificial Neural Network, inspired by the human brain, consists of layers of interconnected nodes (neurons) that process information. A schematic representation of the neural network architecture is shown in Figure 2 (a). The basic structure of the network consists of three layers:

\textbf{(i) Input layer:} The input layer which receives key experimental parameters $x_i$ and passes it to the next layer.\newline
\textbf{(ii) Hidden layer:} The hidden layer to process nonlinear relationships. Depending on the complexity of the problem, the number of hidden layers can be chosen. \cite{Ref26}. For most basic physics laboratory experiments one hidden layer is generally sufficient.\newline
\textbf{(iii) Output layer:} The output layer produces the final result y. \newline
\textbf{(iv) Weights and Bias (ANN parameters):} Additionally, at every artificial neuron (also called Node), the different input values are scaled with the weights ($w_i$), the transfer function is summing the inputs with a bias ($b_h$) as shown in Figure 2(b). The resulting value is processed by the activation function and the result is propagated to the output layer through another bias neuron, $b_o$. The weights ($w_i$) and the bias (b) represent the ANN parameters, which are determined by the training process using suitable algorithms. Different activation functions can be used for the different layers and the common choices are shown in Figure 2 (c). The hidden layers, typically, use sigmoids and/or rectified linear units. At each neuron in the hidden layer, the activation function ($\phi$) transforms the weighted sum of inputs into an output signal as given in the Eq. 8 below:
\begin{equation}
    y =\phi\ (\Sigma w_ix_i + b)
\end{equation}

\begin{figure}[H]
  \centering
  \includegraphics[width=1.0\textwidth]{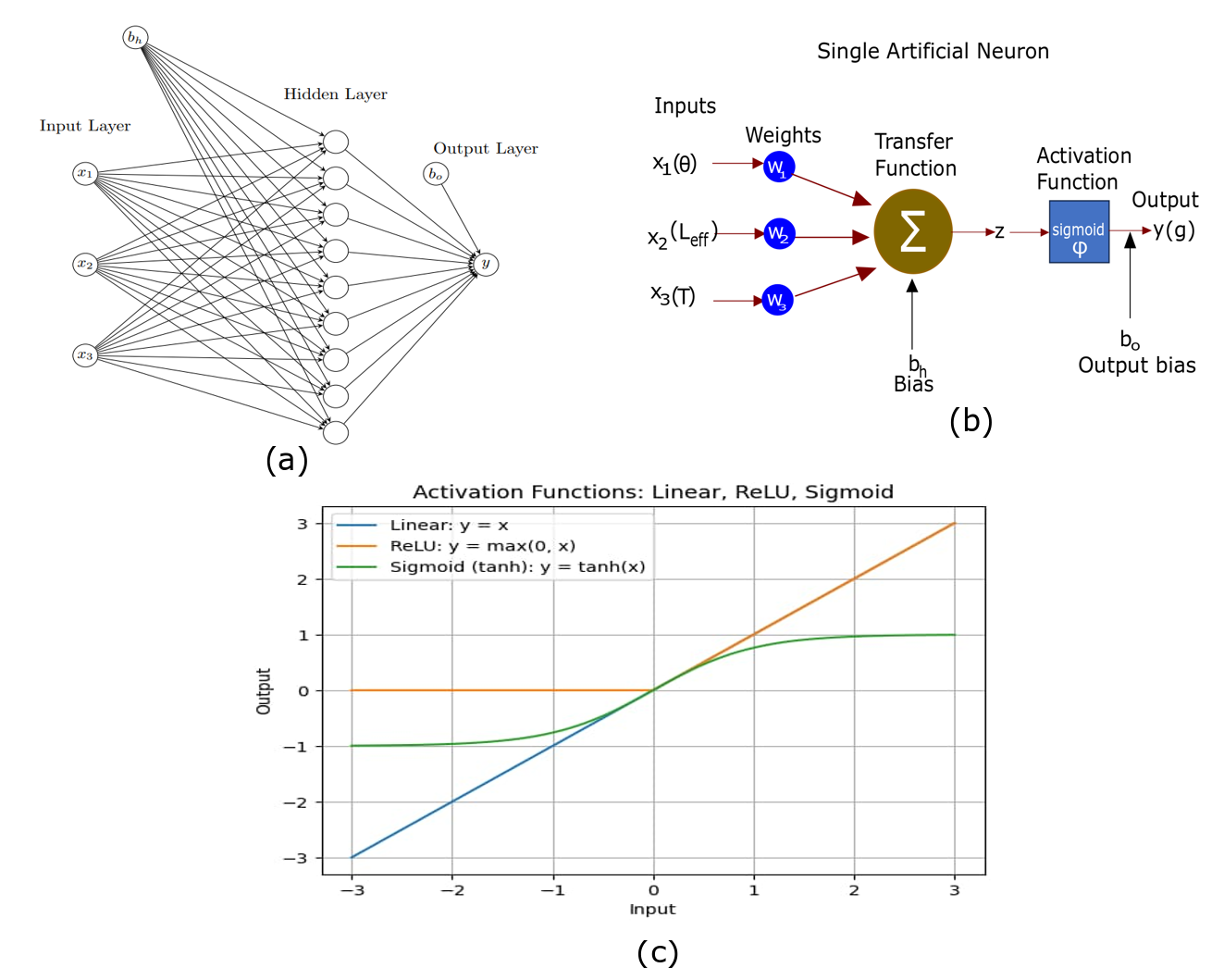}
  \caption{Schematic representation of artificial neural network architecture, neuron model, and common activation functions: (a) a neural network architecture featuring input, hidden, and output layers with biases; (b) a single artificial neuron (node), where the weights (wi) and bias (b) represent the ANN parameters determined during the training process; and (c) typical activation functions for ANNs: linear function, rectified linear unit (ReLU), and sigmoid function.}
  \label{fig:single}
\end{figure}
  
\subsection{ANN Training}
There are a number of training methods available to optimize the weights and biases for a given dataset. Nevertheless, most of these methods use algorithms based on a similar concept: propagation and backpropagation. In this work the ANN training follows the Levenberg–Marquardt algorithm which utilizes the Jacobian matrix (J) and error vector to iteratively update the model parameters \cite{Ref27}. The error vector (e) is defined as difference between actual/true values ($y_{true}$) and predicted outputs ($y_{pred}$):
\begin{equation}
e = y_{true} -y_{pred}		
\end{equation}
The error between the computed values and the provided outputs is computed using two error metrics: (i) Mean Squared Error (MSE) and (ii) Mean absolute error (MAE) which are defined as:
\begin{equation}
MSE = \frac{1}{n} \sum_{i=1}^{n} \left( y_{true,i} - y_{pred,i} \right)^{2}\ \& \ MAE =  \frac{1}{n} \sum_{i=1}^{n} \left| y_{true,i} - y_{pred,i} \right|
\end{equation}
where n represents the number of data points. 
For the compound pendulum experiment the model can be constructed and trained with a basic knowledge of Python and TensorFlow following the steps below:

\textbf{Step 1: Define Neural Network Structure} 
Input Layer: For the compound pendulum experiment, the input layer comprises three experimentally variable parameters as described in the previous section: 
\begin{enumerate}
    \item Initial angular displacement $\theta$
    \item Effective length ($L_{eff}$)
    \item The Time period of oscillation (T), denoted as three neurons $x_1$, $x_2$ and $x_3$.
\end{enumerate}

Hidden Layer: In this model one hidden layer is used comprising nine neurons for optimum performance. In fact the optimization of the number of neurons in the hidden layer can be a good teaching exercise by itself. This layer employs the hyperbolic tangent (tanh) function as the activation function that ensures the outputs to be in the range $-1$ to $+1$, enabling the network to capture intricate patterns within the data.\newline
Output Layer: The output neuron aggregates the processed information from the hidden layer. A single neuron in this layer utilizes a linear activation function, making it suitable for producing continuous numerical predictions.\newline 
\textbf{Step 2: Normalization of Data}
To enhance model stability and speed up convergence, both input and target data are normalized as a preprocessing step. In this case, the min-max normalization method is used to rescale values within the range [0,1]. The expression for normalization is as follows:
\begin{equation}
    x_{norm} = \frac{x - min(x)}{max(x) - min(x)}	
\end{equation}
where, x represents the data value, min(x) and max(x) denote the minimum and maximum values in the dataset, respectively.\newline
\textbf{Step 3: Data Partitioning}
To effectively train and evaluate the model, the input and output dataset is randomly split into three standard subsets:
\begin{itemize}
    \item Training Set: Contains 70\% of the data. It is used to optimize the weights and biases of the network.
    \item Validation Set: Comprises 15\% of the data. This set helps track the model’s performance during training and prevents overfitting by guiding hyperparameter adjustments.
    \item Testing Set: Includes the remaining 15\% of the data. It is used to evaluate the final performance of the trained network.
\end{itemize}
By splitting the dataset in this manner, we ensure that the model is evaluated on unseen data and can generalize well to new inputs.\newline
\textbf{Step-4 : Initializing Weights and Biases}
The parameters of the ANN (weights and biases) are initialized randomly for the first iteration. Small values are preferred to accelerate convergence and prevent activation functions from becoming saturated.\newline
\textbf{Step-5: Computing Predictions}
During each training iteration, the input data is passed through the network layer by layer computing predicted outputs based on the current weight and bias values.\newline
\textbf{Step-6: Loss Calculation for Training Dataset}
The discrepancy between predicted values and actual target values is quantified using the  MSE and MAE as the loss function as described in Eq. 9 and 10. The objective is to minimize this error.\newline
\textbf{Step-7 : Computing the Jacobian Matrix} 
To update ANN parameters, the Jacobian matrix containing the partial derivatives of the loss function with respect to the model parameters ($J = \partial L/ \partial w$) is computed.\newline
\textbf{Step-8: Updating Weights and Biases} Weights and biases are adjusted 
based on the Levenberg-Marquardt algorithm following the update rule: 
$w_{new} = w_{old} + \Delta w$, where
\begin{equation}
\Delta w = -\left(J^{\top}J + \mu I\right)^{-1} J^{\top} e
\end{equation}

Here, $J^\top$ is the transpose of the Jacobian matrix, $\mu$ is a damping factor that controls optimization stability and I is the identity matrix. \newline
\textbf{Step-9 : Adjusting the Damping Factor} 
The damping factor $\mu$ plays a crucial role in controlling the optimization behavior. It is adjusted dynamically based on the performance of the model:
\begin{itemize}
    \item If the loss decreases, $\mu$ is reduced, shifting the method toward the Gauss-Newton approach for faster convergence.
    \item If the loss increases, $\mu$ is increased, making the model rely more on gradient descent to prevent divergence.
\end{itemize} 
This adaptive mechanism ensures a smooth and stable optimization process.\newline
\textbf{Step-10: Monitoring Validation Loss}
To ensure the model generalizes well, the error is calculated for the validation dataset using MSE. This is termed as Validation loss which is continuously monitored. If it ceases to improve over multiple iterations, the training process is adjusted accordingly.\newline
\textbf{Step-11: Training Termination Criteria}
The training process is stopped when one of the following conditions is met:
\begin{itemize}
    \item No improvement in validation loss : If the validation MSE remains stagnant or worsens over several iterations.
    \item Maximum iteration limit reached : If the model completes a predefined number of training cycles.
    \item Convergence achieved: If weight and bias updates become negligible, indicating the model has reached an optimal state.
\end{itemize}
\textbf{Step-12: Final Model Evaluation}
After the completion of training, the model is tested using the testing dataset, ensuring that it can generalize to new data. The obtained outputs are compared with the experimental output data. In the case of compound pendulum, the model performance is evaluated using two error metrics: MSE and MAE. These assessments confirm whether the ANN model is effectively learning patterns and making accurate predictions on unseen inputs\cite{Ref28}, \cite{Ref29}, \cite{Ref30}, \cite{Ref31}. If the performances of the ANN are not sufficient, the ANN structure (e.g., number of hidden layers, number of neurons) or the training algorithms should be updated.
A workflow chart is also depicted in Figure 3 which can be used as a standard laboratory protocol for modeling the compound pendulum problem.

\begin{figure}[H]
  \centering
  \includegraphics[width=0.8\textwidth]{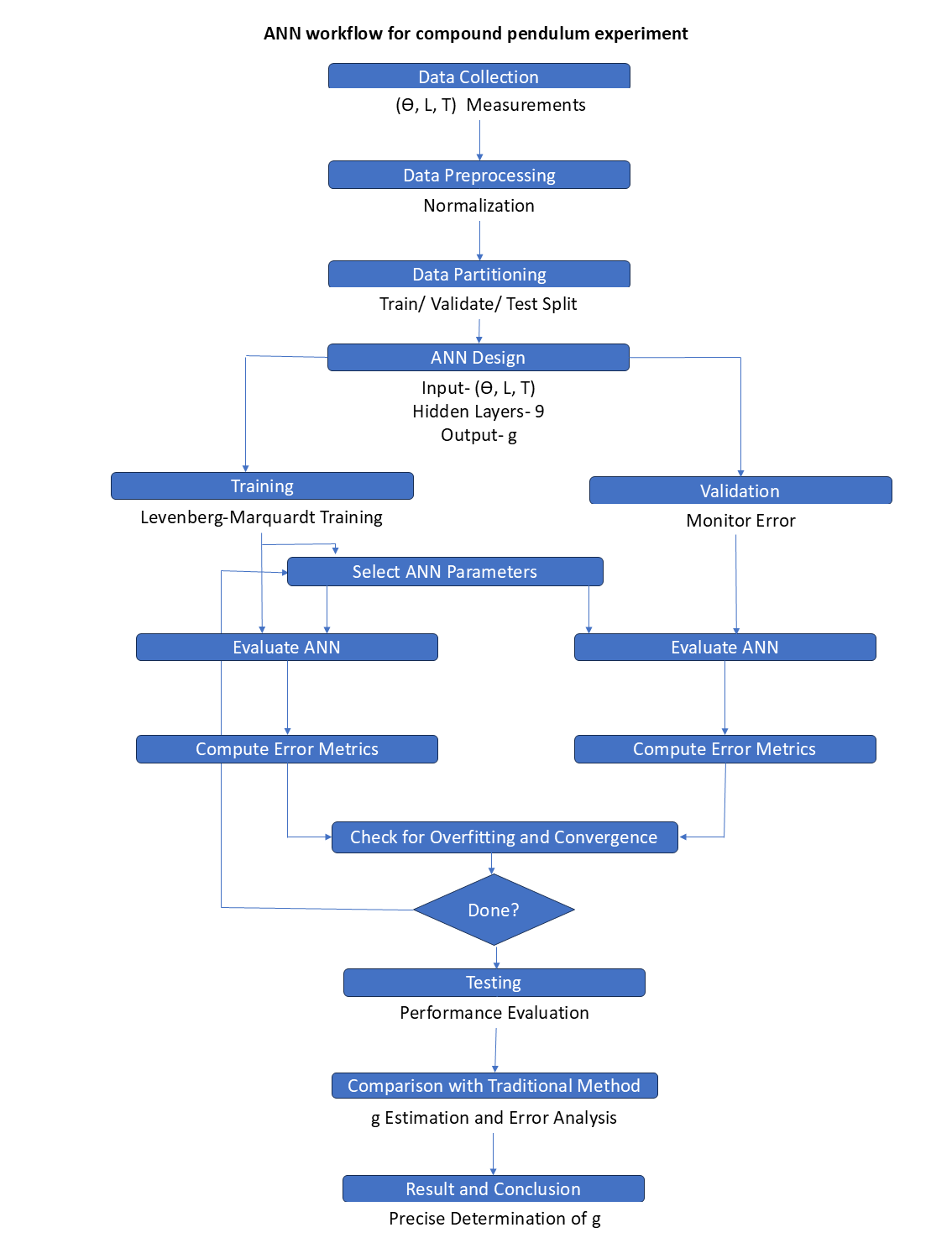}
  \caption{ANN workflow for training the model to compute g and compare with the values obtained by performing the compound pendulum experiment }
  \label{fig:single}
\end{figure}

\newpage
\section{Results and Discussion }
\subsection{Traditional Results:}

Once the experiment is completed, plot of time period (T) versus distance of the pivot point from the centre of the bar for each initial angular displacement. A representative plot is shown in Figure 4, where the variation of T with pivot distance exhibits the expected parabolic behaviour. The effective length of the pendulum ($L_{eff}$) was obtained graphically from the plot as described in Section 2. Using Eq. 7 the value of g was calculated for different values of $L_{eff}$ and initial angular displacement. The experimental data for determining the (g) using a compound pendulum are presented in Table 1. The initial angular displacement was maintained within the small-angle approximation to ensure the validity of the theoretical model.The average value of g was estimated to be 1009.03 cm/s² with a standard deviation of 6.82 cm/s².
\begin{figure}[H]
  \centering
  \includegraphics[width=0.8\textwidth]{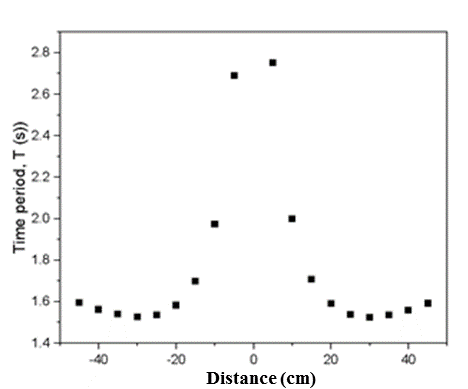}
  \caption{Sample graph for T $\sim$ distance from center of bar at an initial angular displacement of 5° }
  \label{fig:single}
\end{figure}
\begin{center}
\small
\begin{longtable}{cccc}
\caption{Experimental measurements for determining $g$ using a Compound Pendulum.}
\label{tab:I} \\

\toprule
\textbf{Angle (deg)} & \textbf{Equivalent Length, $L$ (cm)} & \textbf{Time Period, $T$ (s)} & $\boldsymbol{g = \frac{4\pi^2 L}{T^2}}$ \textbf{(cm/s$^2$)} \\
\midrule
\endfirsthead

\multicolumn{4}{c}{\small\tablename\ \thetable{} -- \textit{Experimental measurements for determining $g$ using a Compound Pendulum.}} \\[4pt]
\toprule
\textbf{Angle (deg)} & \textbf{Equivalent Length, $L$ (cm)} & \textbf{Time Period, $T$ (s)} & $\boldsymbol{g = \frac{4\pi^2 L}{T^2}}$ \textbf{(cm/s$^2$)} \\
\midrule
\endhead

\midrule
\multicolumn{4}{r}{\small\textit{Continued on next page}} \\
\endfoot

\bottomrule
\endlastfoot

5 & 64.349 & 1.589 & 1005.75 \\
  & 63.978 & 1.583 & 1007.54 \\
  & 63.392 & 1.578 & 1004.65 \\
  & 63.385 & 1.573 & 1010.94 \\
  & 63.063 & 1.569 & 1010.94 \\
  & 62.899 & 1.564 & 1014.77 \\
  & 62.577 & 1.559 & 1016.06 \\
  & 62.065 & 1.554 & 1014.24 \\
  & 61.558 & 1.549 & 1012.46 \\
  & 60.778 & 1.543 & 1007.42 \\
\midrule

4 & 64.349 & 1.586 & 1009.56 \\
  & 63.918 & 1.581 & 1009.15 \\
  & 63.549 & 1.576 & 1009.70 \\
  & 63.036 & 1.572 & 1006.64 \\
  & 62.817 & 1.568 & 1008.28 \\
  & 62.263 & 1.563 & 1005.78 \\
  & 61.997 & 1.559 & 1006.63 \\
  & 61.340 & 1.553 & 1003.68 \\
  & 60.717 & 1.548 & 999.91  \\
  & 60.164 & 1.544 & 995.94  \\
\midrule

3 & 64.315 & 1.578 & 1019.28 \\
  & 63.774 & 1.573 & 1017.15 \\
  & 63.453 & 1.569 & 1017.19 \\
  & 63.064 & 1.564 & 1017.42 \\
  & 62.680 & 1.560 & 1016.43 \\
  & 62.393 & 1.557 & 1015.67 \\
  & 62.188 & 1.553 & 1017.55 \\
  & 61.737 & 1.548 & 1016.71 \\
  & 61.155 & 1.542 & 1014.98 \\
  & 60.909 & 1.537 & 1017.48 \\
\midrule

2 & 64.407 & 1.585 & 1011.74 \\
  & 63.650 & 1.580 & 1006.19 \\
  & 63.107 & 1.575 & 1003.95 \\
  & 62.817 & 1.571 & 1004.44 \\
  & 62.300 & 1.567 & 1001.25 \\
  & 62.028 & 1.564 & 1000.71 \\
  & 61.589 & 1.561 & 997.45  \\
  & 61.007 & 1.555 & 995.67  \\
  & 60.800 & 1.551 & 997.42  \\
  & 60.141 & 1.547 & 991.72  \\
\midrule

1 & 64.165 & 1.583 & 1010.48 \\
  & 63.884 & 1.581 & 1008.62 \\
  & 63.751 & 1.578 & 1010.34 \\
  & 63.761 & 1.574 & 1015.64 \\
  & 63.104 & 1.569 & 1011.60 \\
  & 62.947 & 1.563 & 1016.84 \\
  & 62.475 & 1.558 & 1015.70 \\
  & 61.525 & 1.550 & 1010.60 \\
  & 60.929 & 1.543 & 1009.92 \\
  & 59.859 & 1.536 & 1001.25 \\

\end{longtable}
\end{center}

\subsection{Results of ANN Modeling}
The ANN model, as per its structure design, was trained with 70\% of the experimental dataset by data partitioning. The remaining data were used for validation and testing, 15\% each. The trained neural network parameters consisting of input, hidden and output layers along with the weights and bias values are enlisted in Table 2. Each weight determines how much influence a specific hidden neuron has on the output. Positive weights increase the contribution of the hidden neuron, while negative weights decrease it. A large magnitude weight 2.23750081 indicates that the corresponding hidden neuron strongly influences the output. The bias value -0.0675 is added to the weighted sum of the hidden-layer outputs before applying the activation function at the output layer. Bias helps shift the activation function to fit the target data better, improving model accuracy.

\begin{table*}[h]
    \centering
    \caption{Trained Neural Network Parameters }
    
    \subcaption{(A) Weights (Input-Hidden) and Bias (Input)}
    \resizebox{\textwidth}{!}{
    \begin{tabular}{l c c c c c c c c c}
        \hline
        Variable & HID1 & HID2 & HID3 & HID4 & HID5 & HID6 & HID7 & HID8 & HID9 \\
        \hline
        $\theta$  & 1.52124  & 1.80630  & 0.18349  & 3.68E-4  & 1.81228  & 0.00319  & 1.87E-4  & 0.62452  & 3.34055  \\
        $L_{\text{eff}}$  & 0.74030  & -1.74598  & 1.77497  & 0.88309  & -2.84346  & 0.34463  & -0.87403  & -1.51541  & 1.66830  \\
        $T$  & 1.67771  & 0.86659  & 0.50195  & -0.87028  & 3.23130  & 0.14884  & 0.82275  & 1.54623  & 1.57175  \\
        Bias & -4.15897  & -2.45621  & 2.32752  & -0.67982  & 0.70849  & 0.52462  & -0.65171  & 4.26280  & 1.77722  \\
        \hline
    \end{tabular}
    }

    \vspace{0.5cm}

    \subcaption{(B) Weights (Hidden-Output) and Bias (Output)}
    \resizebox{\textwidth}{!}{
    \begin{tabular}{l c c c c c c c c c}
        \hline
        Variable & HID1 & HID2 & HID3 & HID4 & HID5 & HID6 & HID7 & HID8 & HID9 \\
        \hline
        Output & -0.00070678  & -0.00249705  & 0.00767189  & 2.2335149  & -0.00048397  & 0.26093367  & -2.23750081  & -0.11386300  & -0.000268  \\
        Bias & \multicolumn{9}{c}{-0.0675} \\
        \hline
    \end{tabular}
    }

\end{table*}

The performance of the ANN model was evaluated using two error metrics, namely MSE and MAE, computed for the training, validation, test, and combined (“All”) datasets. Table 3 summarizes the performance of the ANN model for different numbers of hidden nodes ranging from 1 to 30, where each row corresponds to a specific ANN architecture with a given number of neurons in the hidden layer.

Since there is no universal rule for selecting the optimal number of neurons in the hidden layer, a systematic trial-and-error approach was adopted by considering hidden node configurations from 1 to 30. For each configuration, the corresponding ANN model was trained and its predictive performance was evaluated using MSE and MAE. The model yielding the best overall performance, based on the minimum error values and good generalization across training, validation, and test datasets, was selected as the final model for prediction\cite{Ref32}.

The results show a clear dependence of prediction accuracy on the number of hidden nodes. For node configurations such as 2, 3, 5, and 9, both MSE and MAE values are very low across all datasets, indicating excellent predictive capability of the ANN model. Among all tested architectures, the best overall performance was obtained for Node 9, which yielded the lowest overall MAE of 0.0005925 and lowest overall MSE of 0.00150054, demonstrating the highest accuracy and best generalization performance.

Although some configurations such as Nodes 12, 16, 20, 25, and 29 show nearly zero training errors, their validation and test errors are significantly larger, indicating overfitting. In these cases, the model learns the training data extremely well but performs poorly on unseen data. On the other hand, when the number of hidden nodes becomes too large (e.g., 18, 19, 27, and 30), the errors increase substantially, suggesting model instability or excessive complexity.

Conversely, configurations with very few hidden nodes may not adequately capture the nonlinear relationship among the input and output variables, leading to underfitting. Therefore, the ANN results clearly illustrate the importance of selecting an optimal model complexity that balances fitting accuracy and generalization performance.

Based on the obtained results, the ANN model with 9 hidden nodes was selected as the optimum architecture for predicting (g).

The table 4 shows the experimental data for a physical system, likely related to the pendulum motion experiment. Each row represents measurements at a specific angle and associated length of the pendulum. The columns include the following:
\begin{itemize}
    \item \textbf{Angle (°):} The angle at which the pendulum is released. The table shows values between 1° and 5°.
    \item \textbf{Length (cm):} The length of the pendulum measured in centimeters.
    \item \textbf{Times (s):} The measured time period for one complete oscillation of the pendulum in seconds.
    \item \textbf{g (cm/s²):} The calculated acceleration due to gravity using the pendulum's length and the time period of oscillation.
    \begin{itemize}
        \item This value is based on the formula:
        \[
        g = \frac{4\pi^2 L}{T^2}
        \]
        \item Where \textbf{L} is the length of the pendulum and \textbf{T} is the time period.
    \end{itemize}
    \item \textbf{ANN (cm/s²):} The predicted value of (g) from an Artificial Neural Network (ANN) model.
    \begin{itemize}
        \item This value is predicted using a trained model based on the input data (length and time).
    \end{itemize}
\end{itemize}
The values for ‘g’ using compound pendulum and ANN (predicted) are very close to each other, indicating that the ANN model has made accurate predictions that align with the experimental values. 
 The mean value of the ANN-predicted gravitational acceleration obtained from all training, validation, and test datasets was 1009.029858 cm/s², calculated using all predicted outputs from the complete dataset consisting of 34 training values, 8 validation values, and 8 test values (total 50 data points). The optimized model yielded an overall MAE of 0.0005925 cm/s², indicating excellent predictive performance. Therefore, the final ANN-predicted value may be expressed as $g(ANN)=1009.029858 \pm 0.0005925 cm/s^2$, where the error represents the overall MAE of the model. The ANN-predicted mean value is in excellent agreement with the experimentally measured value of gravitational acceleration, 1009.03±6.82 cm/s², where ±6.82 cm/s² represents the standard deviation (SD) of the experimental measurements, confirming the reliability and consistency of the developed neural network model.

\begin{table*}[htbp]
\centering
\caption{Comparison of ANN model performance for different hidden node architectures based on training, validation, test, and overall MAE and MSE values.}
\begin{adjustbox}{max width=\textwidth}
    \begin{tabular}{l cccc | cccc}
        \toprule
        \multirow{2}{*}{\textbf{Nodes}} & \multicolumn{4}{c|}{\textbf{MSE}} & \multicolumn{4}{c}{\textbf{MAE}} \\
        \cmidrule(lr){2-5} \cmidrule(lr){6-9}
        & \textbf{Training} & \textbf{Validation} & \textbf{Test} & \textbf{All} & \textbf{Training} & \textbf{Validation} & \textbf{Test} & \textbf{All} \\
        \midrule
         1 & 0.14606081 & 0.11853278 & 0.07897585 & 0.13324005 & 0.11972002 & 0.09183525 & 0.06504188 & 0.10650995 \\ 
    2 & 0.0008938 & 0.00207514 & 0.00176169 & 0.00131484 & 0.00066134 & 0.00130426 & 0.00131815 & 0.00086929 \\ 
    3 & 0.00083598 & 0.0012409 & 0.00186365 & 0.00113018 & 0.00064088 & 0.00082619 & 0.00127031 & 0.00077124 \\ 
    4 & 0.0917341 & 0.12846464 & 0.59292002 & 0.25418786 & 0.07316727 & 0.10062859 & 0.33454294 & 0.11938119 \\ 
    5 & 0.0008589 & 0.00376868 & 0.00230663 & 0.00190405 & 0.00064954 & 0.00287825 & 0.00153425 & 0.00114768 \\ 
    6 & 0.12092143 & 0.08566678 & 0.32818263 & 0.16837392 & 0.09695062 & 0.06348173 & 0.19739833 & 0.10766723 \\ 
    7 & 0.0004477 & 0.03176589 & 0.01150625 & 0.01351927 & 0.00029376 & 0.01272158 & 0.0089473 & 0.00366678 \\ 
    8 & 0.00378998 & 0.03896136 & 0.00905098 & 0.01630192 & 0.0028721 & 0.02262677 & 0.00707997 & 0.00670611 \\ 
    \textbf{9} & \textbf{0.00018829} & \textbf{0.00172779} & \textbf{0.00330706} & \textbf{0.00150054} & \textbf{0.00014058} & \textbf{0.0011494} & \textbf{0.00195626} & \textbf{0.0005925} \\ 
    10 & 0.00148863 & 0.01162191 & 0.02105585 & 0.00969813 & 0.00115142 & 0.01000584 & 0.01309193 & 0.00447861 \\ 
    11 & 0.02209246 & 0.13825213 & 1.03413497 & 0.41773161 & 0.01864033 & 0.08021528 & 0.44614148 & 0.0968925 \\
    12 & 7.49E-13 & 0.06475514 & 0.02835335 & 0.02827618 & 4.25E-13 & 0.04717697 & 0.02429546 & 0.01143559 \\ 
    13 & 0.03754624 & 0.1925873 & 0.10809361 & 0.09360804 & 0.02874755 & 0.11953536 & 0.07780664 & 0.05112305 \\ 
    14 & 0.02191598 & 0.09792808 & 0.0670702 & 0.05080101 & 0.01740697 & 0.07002668 & 0.04592393 & 0.03038884 \\ 
    15 & 0.05517265 & 0.46009949 & 0.34928413 & 0.2355005 & 0.04178771 & 0.27517355 & 0.21533466 & 0.10689696 \\ 
    16 & 4.63E-13 & 0.11508871 & 0.33711937 & 0.14248923 & 2.84E-13 & 0.08557966 & 0.24440356 & 0.05279732 \\ 
    17 & 0.02486808 & 0.22562632 & 0.30348362 & 0.1526501 & 0.02235981 & 0.16814616 & 0.16242212 & 0.0680956 \\ 
    18 & 0.58600089 & 0.44784148 & 0.70224458 & 0.58694421 & 0.44407192 & 0.4013724 & 0.64551731 & 0.46947126 \\ 
    19 & 0.92098877 & 0.93127665 & 1.23129058 & 0.97883924 & 0.80869679 & 0.80411174 & 1.08048237 & 0.85144888 \\ 
    20 & 2.04E-07 & 1.13864457 & 1.44266106 & 0.73514977 & 1.56E-07 & 0.73536754 & 0.93835261 & 0.26779533 \\ 
    21 & 0.26099689 & 1.3978526 & 1.26729065 & 0.78480824 & 0.17433178 & 1.13163167 & 1.03578714 & 0.46533262 \\ 
    22 & 0.06731807 & 0.81923151 & 0.33042596 & 0.35767726 & 0.04928146 & 0.72023916 & 0.2794118 & 0.19345555 \\ 
    23 & 0.002389 & 0.41644934 & 0.41079292 & 0.23399321 & 0.00187841 & 0.34487388 & 0.30844553 & 0.10580842 \\ 
    24 & 2.19789249 & 2.68581961 & 2.21438637 & 2.28552812 & 1.5696436 & 2.27325991 & 1.76746859 & 1.71387421 \\ 
    25 & 4.36E-14 & 0.74067125 & 1.81080897 & 0.78257248 & 1.67E-14 & 0.69103308 & 1.34977135 & 0.32652871 \\ 
    26 & 0.00325085 & 0.29868209 & 0.92719814 & 0.38965673 & 0.00221918 & 0.23030641 & 0.73775136 & 0.15639829 \\ 
    27 & 0.0362579 & 3.00925711 & 1.05259361 & 1.27556537 & 0.02636468 & 2.63371236 & 0.91274196 & 0.58536067 \\ 
    28 & 0.18278312 & 0.74749544 & 1.10754337 & 0.5553223 & 0.13947942 & 0.56378219 & 0.82969653 & 0.3178026 \\ 
    29 & 1.21E-08 & 1.3305028 & 1.50626845 & 0.80389873 & 8.97E-09 & 0.94801074 & 1.13351828 & 0.33304465 \\ 
    30 & 0.39132282 & 2.88634485 & 1.2313495 & 1.29602628 & 0.32292021 & 2.6574005 & 0.94349315 & 0.79572873 \\
        \bottomrule
    \end{tabular}
\end{adjustbox}
\end{table*}

\begin{figure}[H]
  \centering
  \includegraphics[width=1.0\textwidth]{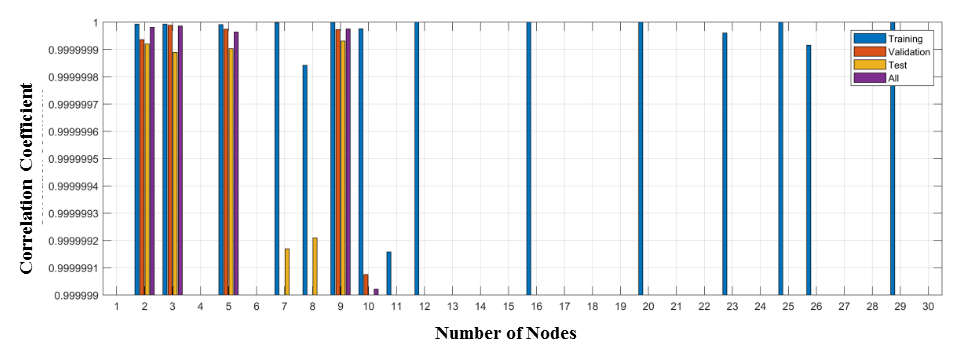}
  \caption{Correlation Coefficient vs. Number of Nodes for Training, Validation, Test, and Overall Datasets in ANN.}
  \label{fig:single}
\end{figure}

The correlation coefficient (r) measures the strength of the relationship between the input variables and is computed using normalized covariance which falls in a range between -1 to +1. 
\begin{equation}
    r = \frac{\sum(y_i - \bar{y})(\hat{y}_i - \bar{\hat{y}})}{\sqrt{\sum(y_i - \bar{y})^2 \sum(\hat{y}_i - \bar{\hat{y}})^2}}
\end{equation}

A high value of r ($\approx$ 1) signifies that the model predictions are highly aligned with the actual outputs. Figure 5 shows the correlation coefficient (r-values) for the training, validation, test, and overall (all) datasets as a function of the number of nodes. The x-axis represents the number of nodes (from 1 to 30), and the y-axis represents the correlation coefficient, with a very fine range between 0.999991 and 1. The correlation coefficients for all datasets (training, validation, test, and overall) are very close to 1 signifying that the model predictions are highly aligned with the actual outputs. Nodes 1-5, 8-10, and 23-30 show correlation coefficients nearly equal to 1 across the training, validation, testing, and overall datasets indicating optimal performance at these specific node configurations. Nodes 6, 7, and 11 show a decline in correlation coefficients, especially in the validation and test datasets implying reduced generalization, possibly due to underfitting or overfitting. The absence of correlation coefficient bars at certain nodes suggests that the ANN did not converge or performed poorly at these configurations. Training Dataset (blue bars) shows consistently high R-values across most nodes, reflecting the model’s ability to fit the training data effectively. For Validation Dataset (yellow bars) the R-values are lower at certain nodes (e.g., 6, 7), suggesting a reduction in generalization. Test Dataset (red bars) exhibits similar behavior to the validation dataset, further confirming that certain nodes (6, 7, and 11) impact the model’s predictive ability on unseen data. All Dataset (purple bars) maintains a high R-value across optimal nodes, reaffirming overall model reliability. This provides an instructive example of how correlation metrics should be interpreted in conjunction with validation performance rather than in isolation.

The training performance of the ANN model is illustrated in Figure 6, where the MSE is plotted on a logarithmic scale against the number of training iterations (epochs) for the training, validation, and test datasets. A rapid decrease in MSE is observed during the initial stage of training (epochs 0–20) for all three curves, indicating fast learning as the network adjusts its weights. As training progresses, the training error (blue curve) continues to decrease gradually, while the validation error (green curve) and test error (red curve) remain relatively stable. The validation MSE reaches its minimum value of $2.9853 \times 10^{-6}$
 at epoch 127, as indicated by the green circle in Figure 6. This point represents the optimum balance between model learning and generalization performance. Beyond epoch 127, the validation error begins to increase continuously for several successive epochs, which is a clear indication of the onset of overfitting. According to the early stopping criterion, training was therefore terminated at this stage to prevent deterioration of model performance on unseen data. Although the training process continued up to 133 epochs, the network parameters corresponding to epoch 127 were retained as the best model. The close agreement between the validation and test error curves throughout the training process confirms that the developed ANN model generalizes well and provides reliable predictions without significant underfitting or overfitting.

\begin{figure}[H]
    \centering
    \includegraphics[width=1.0\textwidth]{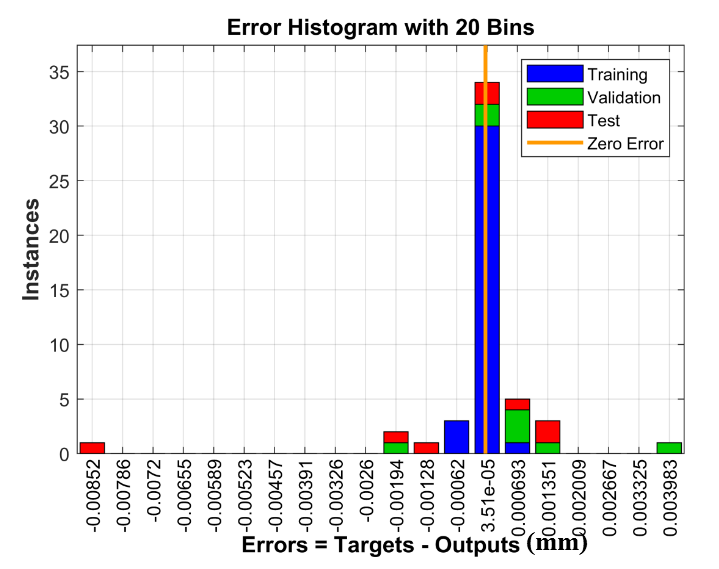} 
    \caption{The performance of the artificial neural network model for the ‘g’ value based on the MSE across training, validation, and testing.}
    \label{fig:sample}
\end{figure}

\subsection{Comparison of results}
In traditional experiments, uncertainties arise from physical and instrumental limitations such as measurement fluctuations, timing errors, and environmental effects, whereas in ANN modeling the prediction error arises from the model’s ability to learn and approximate relationships present in the dataset. The experimentally estimated and ANN-predicted values of g for the compound pendulum experiment corresponding to the training, validation, and test datasets are presented in Tables 4, 5, and 6, respectively. The associated absolute error and absolute percentage error are also reported. The close agreement between the experimentally determined and ANN-predicted values of g indicates that the ANN model successfully learned and reproduced the relationship between the measured input parameters (initial angular displacement, effective length, and time period) and the corresponding values of g. However, the ANN prediction error should be interpreted as a measure of model-fitting performance rather than as an improvement in the intrinsic experimental precision of the measurement. The model performance and prediction consistency were further analyzed using regression analysis and error histograms for the training, validation, and test datasets.

\begin{table}
    \centering
    \caption{Experimental and ANN-Predicted Values of g, Absolute Error, and Absolute Percentage Error for the Training Dataset}
    \renewcommand{\arraystretch}{1.5}
    \setlength{\tabcolsep}{6pt}
    \resizebox{\textwidth}{!}{
    \begin{tabular}{c c c c c c c}
       \hline
        Angle (deg) & Length $L$ (cm) & Time Period $T$ (s) & $g_{exp}$ (cm/s\textsuperscript{2}) & $g_{ANN}$ (cm/s\textsuperscript{2}) & Absolute Error ($\times10^{-4}$ cm/s\textsuperscript{2}) & Absolute Percentile Error ($\times10^{-7}$) \\ \hline

        5  & 64.34925  & 1.589 & 1005.753515 & 1005.753547 & 0.3159 & 0.3141 \\
           & 63.392    & 1.578 & 1004.653543 & 1004.653442 & 1.0051 & 1.0004 \\
           & 63.385    & 1.573 & 1010.938913 & 1010.938870 & 0.4296 & 0.4250 \\
        \midrule
        4  & 64.34925  & 1.586 & 1009.561982 & 1009.561799 & 1.8313 & 1.8139 \\
           & 63.0355   & 1.572 & 1006.644170 & 1006.644527 & 3.5653 & 3.5418 \\
           & 62.81725  & 1.568 & 1008.283517 & 1008.283574 & 0.5714 & 0.5667 \\
        \midrule
        3  & 64.315    & 1.578 & 1019.281497 & 1019.281652 & 1.5492 & 1.5199 \\
           & 63.77425  & 1.573 & 1017.147132 & 1017.147195 & 0.6336 & 0.6229 \\
           & 63.45275  & 1.569 & 1017.186121 & 1017.185832 & 2.8895 & 2.8407 \\
        \midrule
        2  & 64.40675  & 1.585 & 1011.739523 & 1011.739580 & 0.5709 & 0.5642 \\
           & 63.65025  & 1.580 & 1006.194187 & 1006.193957 & 2.3046 & 2.2904 \\
           & 63.107    & 1.575 & 1003.950460 & 1003.950595 & 1.3541 & 1.3487 \\
        \midrule
        1  & 64.1645   & 1.583 & 1010.482621 & 1010.482428 & 1.9296 & 1.9096 \\
           & 63.88425  & 1.581 & 1008.616166 & 1008.616593 & 4.2715 & 4.2350 \\
           & 63.751    & 1.578 & 1010.343072 & 1010.342723 & 3.4865 & 3.4508 \\
        \hline
    \end{tabular}
    }
\end{table}

\begin{table}
    \centering
    \caption{Experimental and ANN-Predicted Values of g, Absolute Error, and Absolute Percentage Error for the Validation Dataset}
    \renewcommand{\arraystretch}{1.5}
    \setlength{\tabcolsep}{4pt}
    \resizebox{\linewidth}{!}{
    \begin{tabular}{c c c c c c c c}
        \hline
        Sl No. & Angle (deg) & Length (cm) & Time (s) & $g$ (cm/s\textsuperscript{2}) & $g_{ANN}$ (cm/s\textsuperscript{2}) & Absolute Error ($\times10^{-3}$ cm/s\textsuperscript{2}) & Absolute Percentile Error ($\times10^{-4}$) \\
        \hline
        1  & 5  & 62.899    & 1.564 & 1014.766467 & 1014.768090 & 1.6230 & 1.5994 \\
        2  & 5  & 62.577    & 1.559 & 1016.057703 & 1016.057604 & 0.0990 & 0.0974 \\
        3  & 4  & 63.91825  & 1.581 & 1009.152964 & 1009.152309 & 0.6550 & 0.6491 \\
        4  & 4  & 63.54925  & 1.576 & 1009.703513 & 1009.702876 & 0.6370 & 0.6309 \\
        5  & 3  & 63.06375  & 1.564 & 1017.424423 & 1017.423150 & 1.2730 & 1.2512 \\
        6  & 3  & 62.18775  & 1.553 & 1017.554775 & 1017.550463 & 4.3120 & 4.2376 \\
        7  & 2  & 62.028    & 1.564 & 1000.714390 & 1000.714618 & 0.2280 & 0.2278 \\
        8  & 1  & 59.859    & 1.536 & 1001.250832 & 1001.250463 & 0.3690 & 0.3685 \\
        \hline
    \end{tabular}
    }
\end{table}

\begin{table}
    \centering
    \caption{Experimental and ANN-Predicted Values of g, Absolute Error, and Absolute Percentage Error for the Test Dataset}
    \renewcommand{\arraystretch}{1.5}
    \setlength{\tabcolsep}{4pt}
    \resizebox{\linewidth}{!}{
    \begin{tabular}{c c c c c c c c}
        \hline
        Sl No. & Angle (deg) & Length (cm) & Time (s) & $g$ (cm/s\textsuperscript{2}) & $g_{ANN}$ (cm/s\textsuperscript{2}) & Absolute Error ($\times10^{-3}$ cm/s\textsuperscript{2}) & Absolute Percentile Error ($\times10^{-4}$) \\
        \hline
        1  & 5  & 63.97775  & 1.583 & 1007.541600 & 1007.540500 & 1.1000 & 1.0918 \\
        2  & 4  & 60.1635   & 1.544 & 995.942700  & 995.942100  & 0.6000 & 0.6024 \\
        3  & 3  & 61.73675  & 1.548 & 1016.711500 & 1016.710400 & 1.1000 & 1.0819 \\
        4  & 4  & 61.15475  & 1.542 & 1014.988400 & 1014.989400 & 1.0000 & 0.9852 \\
        5  & 2  & 62.81725  & 1.571 & 1004.436000 & 1004.436300 & 0.3000 & 0.2987 \\
        6  & 2  & 61.00725  & 1.555 & 995.672400  & 995.672600  & 0.2000 & 0.2009 \\
        7  & 1  & 63.104    & 1.569 & 1011.595400 & 1011.597500 & 2.1000 & 2.0759 \\
        8  & 1  & 61.5245   & 1.550 & 1010.602900 & 1010.604400 & 1.5000 & 1.4843 \\
        \hline
    \end{tabular}
    }
\end{table}

\begin{figure}[H]
  \centering
  \includegraphics[width=1.0\textwidth]{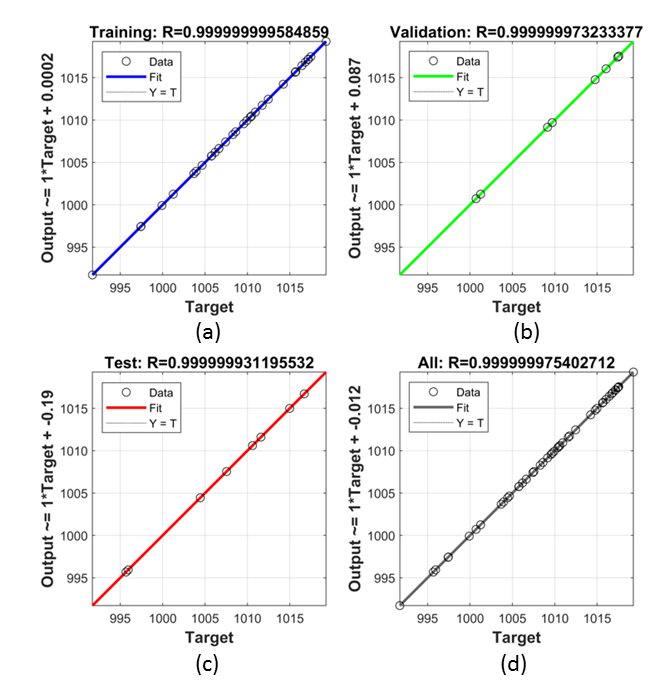}
  \caption{Regression plots illustrating the comparison between ANN predictions and experimental measurements for (a) Training, (b) Validation, (c) Test, and (d) All datasets. }
  \label{fig:single}
\end{figure}
\newpage
A linear regression analysis was performed comparing the ANN-predicted values with experimental values and are presented in Figure 7 for the training, validation, and testing datasets. The strong linear correlation between ANN predicted and experimental values of g and the correlation coefficient (R ) values close to unity, indicate that the ANN model successfully captures the relationship between the input parameters and g. While the ANN predictions exhibit very small deviations, it is important to interpret these results within the context of experimental uncertainty and model assumptions.

Errors in ANN model are defined as the difference between targets (true values) and outputs (predicted values). The distribution of prediction errors in the ANN model is studied graphically through an error histogram for Training, Validation and Test datasets as shown in Fig. 7. The error distribution is represented across 20 bins. The orange line at the center represents the exact zero-error point. The central peak, located near zero error mark, indicates that the ANN model has a high prediction accuracy, with the majority of the instances (for training, validation, and test) concentrated around the zero-error mark. This suggests that the model has successfully minimized the error during training and generalizes well to unseen data. The training dataset (Blue) exhibits the most significant concentration of instances near zero error, demonstrating that the model has effectively learned the patterns in the training data. A smaller proportion of errors are associated with the validation data set (Green), but they are also close to zero error, reflecting good generalization. The test data set (Red) also shows a small error spread, indicating that the model maintains its accuracy on unseen data. The symmetry and proximity of most errors around this line demonstrate the model’s precision. A few instances on the left and right tails of the histogram indicate small deviations from zero error. However, these errors are minimal and do not significantly impact the model’s overall performance.

 The ANN-predicted value shows better precision compared with the experimental result because the neural network learns the nonlinear relationship between the input parameters and gravitational acceleration from the available dataset, thereby reducing the effect of random experimental fluctuations and measurement noise. While the experimental value (1009.03±6.82) cm/s² contains uncertainties arising from instrumental limitations, human error, and environmental variations, the ANN model provides a highly consistent prediction with very low error. In addition, once trained, the ANN model can rapidly estimate g for new combinations of input parameters ($x_1, x_2, x_3$) without performing repeated physical experiments. This makes the model particularly useful for exploring a wider range of conditions, optimizing parameters, and predicting outcomes in situations where experimental measurements may be time-consuming, difficult, or not easily achievable.

\begin{figure}[H]
    \centering
    \includegraphics[width=1.0\textwidth]{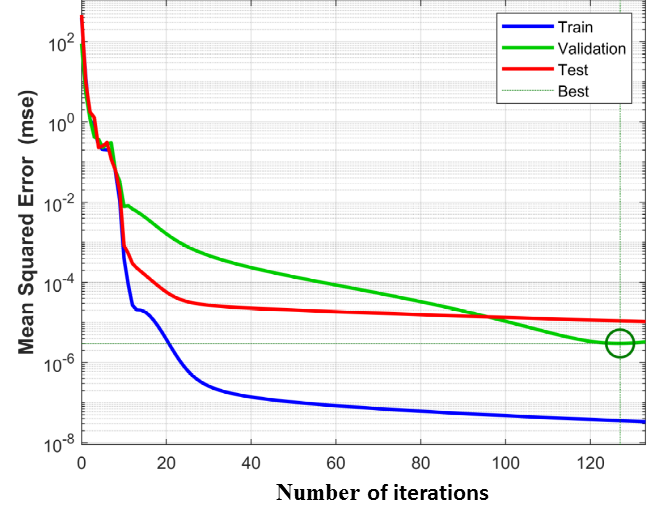} 
    \caption{The histogram plot illustrates the distribution of error values for ’g’ during the training, validation, and testing phases of the artificial neural network model.}
    \label{fig:sample}
\end{figure}

\section{Pedagogical Approach} 
From a pedagogical perspective, the comparison between experimentally determined and ANN-predicted values of g provides an opportunity for students to get an exposure to multiple approaches of data analysis. The ANN model does not replace the traditional method but serves as a complementary tool that enables students to explore concepts such as model fitting, generalisation, and validation. The variation of error with network complexity, the analysis of correlation coefficients and the study of training dynamics together provide a coherent framework for understanding both experimental uncertainty and data-driven modelling. This integrated approach encourages students to critically evaluate results rather than relying solely on numerical agreement.
\subsection{Classroom implementation}
The primary objective of this work is to introduce Artificial Neural Networks (ANNs) into the undergraduate physics laboratory through hybrid experiments in a structured approach by blending theoretical concepts, experimental investigation and data-driven modeling. The proposed activity may be incorporated for the second/third-year undergraduate laboratory course on oscillatory systems or experimental methods in physics or engineering programmes. However, the implementation of such an activity requires careful consideration. Students may have varying levels of familiarity with programming and machine learning concepts, which may necessitate additional instructional support. Providing structured guidelines and simplified code frameworks can help ensure that the focus remains on conceptual understanding rather than technical complexity. It can be designed as an activity of 2-3  laboratory sessions, allowing sufficient time for both experimental data acquisition and computational analysis.
The activity is proposed to be structured in three stages: 
\begin{enumerate}
    \item First stage: Students perform the compound pendulum experiment and measure the time period for different effective lengths and small angular displacements. They record and organise the data systematically.
    \item Second stage: Students analyse the data using conventional methods to determine (g), including estimation of uncertainties. This step reinforces fundamental concepts such as data fitting, graphical analysis, and error propagation.
    \item Third stage: Students are introduced to an ANN-based approach for modelling the same dataset as a second layer of analysis. A portion of the experimental data is used to train the model, while the remaining data are used for validation. Students then compare the ANN predictions with the results obtained from the analytical method. This comparison forms the basis for discussion on model reliability, generalisation, and limitations.
\end{enumerate}   
The activity is designed to promote collaborative and inquiry-based learning. Students typically work in small groups, engaging in both experimental and computational tasks. The instructor’s role is to guide interpretation rather than provide direct solutions, thereby encouraging independent reasoning. Student understanding may be assessed through laboratory reports including error comparison, short computational assignments, and oral discussions on limitations of both methods. Particular emphasis is placed on the interpretation of results, rather than on numerical agreement alone.

\subsection{Educational outcomes}
Completion of this hybrid experiment would enable the students to understand the following questions:
\begin{enumerate}
 \item WHAT are the practical limitations of theoretical models in real world experiments
\item HOW to apply ANN for scientific modeling of complex, nonlinear relationships without explicit equations 
\item WHY ANN gives lower uncertainty and better consistency and finally,
\item WHAT is the difference between theory-driven and data-driven approaches
\end{enumerate}

Overall, this approach transforms a standard laboratory exercise into an interdisciplinary learning experience by introducing students to machine learning while simultaneously strengthening their understanding of experimental physics. One of the key educational outcomes is an improved understanding of experimental uncertainty. While the conventional approach focuses on traditional error analysis, the ANN framework enables students to explore how models learn from experimental data, thereby providing a complementary perspective on data interpretation.

The activity also introduces important concepts such as model selection and generalisation. By examining how prediction errors vary with the number of hidden nodes, students gain insight into underfitting and overfitting, helping them understand that increasing model complexity does not necessarily improve predictive performance. This idea is equally important in both experimental physics and data science.

Another significant outcome is the development of critical thinking skills. By comparing theory-driven and data-driven approaches, students learn to evaluate the role of assumptions, approximations, and data quality in scientific analysis.

\subsection{Limitations of ANN-Based Modeling in Experimental Physics}

The ANN-based analysis adopted in this work also provides an opportunity to discuss the practical limitations of data-driven modeling in experimental physics. Unlike analytical methods derived from established physical principles, a conventional ANN does not explicitly incorporate governing physical laws and may therefore produce predictions that are mathematically accurate but not necessarily physically interpretable. Moreover, the performance of the network depends strongly on the quality, quantity, and variability of the training data. If model complexity is not properly controlled, the ANN may overfit experimental noise and reproduce random fluctuations instead of the underlying physical behaviour. In addition, ANN models do not directly provide physically meaningful parameters in the same way as conventional analytical fitting techniques.

These limitations form an important pedagogical aspect of the laboratory exercise. Rather than treating ANN methods as replacements for traditional experimental analysis, the activity encourages students to critically examine the advantages and limitations of both analytical and data-driven methodologies. This helps students appreciate the importance of physical interpretability, uncertainty analysis, and the incorporation of physical constraints in modern computational modeling.

The discussion also highlights that ANN models cannot fully replace analytical models grounded in physical theory. Although neural networks are powerful tools for pattern recognition and prediction, they do not inherently explain the physics governing a system. Consequently, combining data-driven techniques with physics-based reasoning is essential for developing reliable and scientifically meaningful models in experimental research.

\section{Conclusion}

This work demonstrates the implementation of an ANN framework in a traditional compound pendulum experiment for determining the value of g, in an undergraduate physics laboratory. The experimentally determined value, $g(exp) =1009.03\ \pm \ 6.82 cm/s^2$, showed close agreement with the ANN-predicted value, $g(ANN)=1009.029858 \pm 0.0005925\ cm/s^2$. The ANN model successfully learned the relationship between the experimentally measured pendulum parameters and the corresponding values of g within the available dataset. To validate the ANN model, a systematic analysis was carried out using different hidden-node architectures based on training, validation, test, and overall MAE and MSE values. The model performance was further examined through correlation coefficient analysis, regression plots, MSE performance curves, and comparison of experimental and ANN-predicted values for the training, validation, and test datasets. In addition, histogram plots of the prediction errors during the training, validation, and testing phases were analyzed to examine the distribution of errors and the consistency of model predictions. The close agreement between experimental and predicted values, together with high correlation coefficients and low prediction errors, confirms the reliability of the ANN model in reproducing the trends present in the experimental data. However, the ANN prediction error should not be interpreted as an improvement in the intrinsic experimental precision, since the network is trained using experimental data governed by the known analytical relation of the compound pendulum.
The primary significance of this work lies in its pedagogical contribution. The proposed framework provides a practical method for integrating machine learning with conventional experimental physics, enabling students to explore concepts such as regression, training, validation, overfitting, generalization, and data-driven analysis alongside traditional experimental methods and error estimation. This interdisciplinary approach can be extended to other undergraduate laboratory experiments, providing a useful pathway for modernizing physics laboratory education.

\section*{Acknowledgment}
 We sincerely acknowledge the invaluable assistance of Mr. Susant Ku Parida and Mr. Debasis Das, Technicians, School of Physical Sciences, NISER Bhubaneswar, in the data collection for the compound pendulum experiment aimed at measuring (g). Their efforts in conducting precise measurements and ensuring the accuracy of the experimental data have significantly contributed to the success of this study. We deeply appreciate their dedication and support.

\bibliographystyle{unsrt}
\bibliography{Reference}

\end{document}